\documentclass[9pt, sigconf]{acmart}
\usepackage[nolist]{acronym}
\usepackage{cleveref}
\usepackage{subcaption}
\usepackage{hyperref}

\AtBeginDocument{%
  }

\setcopyright{acmcopyright}
\copyrightyear{2024} 
\acmYear{2024} 
\setcopyright{acmlicensed}
\acmConference[E-Energy '24]{The 15th ACM International Conference on Future and Sustainable Energy Systems}{June 4--7, 2024}{Singapore, Singapore}
\acmBooktitle{The 15th ACM International Conference on Future and Sustainable Energy Systems (E-Energy '24), June 4--7, 2024, Singapore, Singapore}
\acmDOI{10.1145/3632775.3661946}
\acmISBN{979-8-4007-0480-2/24/06}

\begin{document}

\title{Benchmarking Aggregation-Dis\-aggregation Pipelines for Smart Charging of Electric Vehicles}

\author{Leo Strobel}
\orcid{0000-0002-1069-315X}
\affiliation{%
  \institution{University of Würzburg}
  \streetaddress{Am Hubland}
  \city{Würzburg}
  \country{Germany}
  \postcode{97074}
}
\email{leo.strobel@uni-wuerzburg.de}

\author{Marco Pruckner}
\affiliation{%
  \institution{University of Würzburg}
  \streetaddress{Am Hubland}
  \city{Würzburg}
  \country{Germany}
  \postcode{97074}
}
\email{marco.pruckner@uni-wuerzburg.de}

\renewcommand{\shortauthors}{Strobel et al.}

\begin{abstract}
    As the global energy landscape shifts towards renewable energy and the electrification of the transport and heating sectors,
    national energy systems will include more controllable prosumers.
    Many future scenarios contain millions of such prosumers with individualistic behavior.
    This poses a problem for energy system modelers.
    Memory and runtime limitations often make it impossible to model each prosumer individually.
    In these cases, it is necessary to model the prosumers with representatives or in aggregated form.
    Existing literature offers various aggregation methods,
    each with strengths, drawbacks, and an inherent modeling error.
    It is difficult to evaluate which of these methods perform best.
    Each paper presenting a new aggregation method usually includes a performance evaluation.
    However, what is missing is a direct comparison on the same benchmark,
    preferably by a neutral third party that is not associated with any of the compared methods.
    This paper addresses this gap by introducing a benchmark to evaluate the end-to-end performance of multiple 
    aggregation-disaggregation pipelines, specifically focusing on \acp{EV}.
    Our study assesses the performance of the common representative profile (REP) approach,
    four different versions of the \ac{VB} approach, the \ac{FO} method, and the \ac{DFO} method.
    The results show that each method has a clear use case.
    Depending on the price signal, additional median charging costs of 2\%-50\% are incurred using an aggregation method,
    compared to the optimal charging costs (i.e., the charging costs resulting from optimizing the \acp{EV} directly, without aggregation).
    The representative profile approach results in the lowest additional costs (2\%-20\%),
    while the \ac{FO} and \ac{DFO} methods allow for error-free disaggregation,
    which is advantageous in real-world use cases.
\end{abstract}




\maketitle

\section{Introduction}

\Acp{EV} are rapidly gaining market share, and with ongoing widespread
governmental support, it seems certain that they will replace internal combustion engine
vehicles as the standard propulsion type in the coming decades \cite{internationalenergyagencyGlobalEVOutlook2023}.
Consequently, we can expect \ac{EV} numbers in the hundreds of millions globally in the medium term.
The risk that uncontrolled charging of such fleets poses for the stability of the electric grid is well known and has been analyzed in several studies
\cite{strobelJointAnalysisRegional2022,doluweeraScenariobasedStudyImpacts2020,pieltainfernandezAssessmentImpactPlugin2011,salahImpactElectricVehicles2015}.
Conversely, the possible benefits of smart charging of such fleets are also well established
\cite{strobelJointAnalysisRegional2022,doluweeraScenariobasedStudyImpacts2020,apostolaki-iosifidouElectricVehicleCharge2020,thingvadEconomicValueElectric2019}.

These two possible extremes show that the impact of \acp{EV} on the power system is not straightforward and, at the same time, highly significant.
Therefore, it is clear that \acp{EV} must be integrated into future energy system models
and that they must be modeled as active components that respond to the control decisions of other energy system actors.
However, integrating millions of active components into classical unit commitment models \cite{padhyUnitCommitmentaBibliographical2004a} presents a challenge for modelers.
In most cases,
modeling every \ac{EV} individually is not feasible due to memory and runtime restrictions.
Even if technically possible in some cases,
it is undesirable to spend the vast majority of the computational budget on just one component of the energy system,
especially in a modeling field where model complexity is already a limiting factor \cite{padhyUnitCommitmentaBibliographical2004a}.

If modeling each \ac{EV} individually is not an option, two widely used alternatives remain.
Firstly, the use of representative vehicles or profiles.
These profiles are modeled in the same manner as individual vehicles, with the only difference being that the resulting load is scaled up to represent a larger fleet.
This approach is straightforward and generally leads to realistic results,
but it comes with the disadvantage that the number of necessary profiles is often quite large.
Different studies arrive at different numbers of needed profiles:
Strobel et al. found 100 profiles to be acceptable \cite{strobelJointAnalysisRegional2022},
Schlund et al. recommend 1000 profiles \cite{schlundElectricVehicleCharging2021},
and Müssel et al. recommend between 5000 to 12,000 profiles \cite{muesselAccurateScalableRepresentation2023a}.
The exact number of profiles necessary will depend on the research question and the desired temporal resolution of the result.
Generally, optimization is feasible in a reasonable time with all the recommended numbers.
However, when compared to the approximately 3000 power plants with more than 100 MW in the European energy system \cite{entso-ePowerStatistics2022},
it becomes clear that this approach still demands a large portion of the computational budget.
This problem is exacerbated when spatially resolved results are required,
for example in grid development planning \cite{KurzstudieElektromobilitatModellierung2019}.
In that case, each spatial node will require the minimum recommended profiles,
quickly leading to intractable profile counts.

The second alternative is to aggregate the \ac{EV} to abstract objects that encode the combined flexibility of the fleet.
The energy system model is then optimized with these aggregated objects,
and the result is disaggregated back to individual \acp{EV}.
Several possible aggregation methods have been proposed in the literature.
The majority of these can be grouped under the virtual battery/storage umbrella
\cite{pertlEquivalentTimeVariantStorage2019,zhouFormingDispatchableRegion2021,gunkelPassiveActiveFlexibility2020,vayaSmartChargingPlugin2012,brinkelSchedulingElectricVehicle2022,haoCharacterizingFlexibilityAggregation2014}.
These methods are characterized by modeling the \ac{EV} fleet as one or several energy storages
that usually have the parameters virtual power, min/max \ac{SOC}, and energy loss due to the mobility demand of the \acp{EV}.
Generally, these models overestimate the flexibility of the \ac{EV} fleet,
for reasons we will discuss in \Cref{sec:vb},
and it is not necessarily possible to disaggregate the electric load precisely to the \acp{EV}.
Methods that are perfectly disaggregateable include the FlexObject approach \cite{siksnysAggregatingDisaggregatingFlexibility2015a,pedersenModelingManagingEnergy2018a,neupaneGenerationEvaluationFlexOffers2017a},
the Dependency-based FlexOffers \cite{siksnysDependencybasedFlexOffersScalable2016},
and the FlexAbility approach \cite{schlundFlexAbilityModelingMaximizing2020}.
The guarantee of feasible disaggregation is highly desirable.

If we look at the entire aggregation, optimization/simulation, and disaggregation pipeline,
all aggregation approaches we found in the literature will lead to information losses and,
consequently, underestimate the value of smart charging.
However, how large these errors are compared to each other needs to be determined.
A benchmark that evaluates each method on equal grounds is necessary.
In this paper, we present such a benchmark and use it to determine which of the highlighted methods works best
in terms of runtime and the additional charging costs introduced by aggregation.
The benchmarking setup includes a driving profile generator based on the \ac{MID} dataset.
Additionally, the benchmark provides an unaggregated optimization of \ac{EV} charging and a simulation of uncontrolled charging
as upper and lower reference points for possible charging costs.
The performance is evaluated based on three different price signals.
To run the benchmark, we reimplemented each method in Python and Rust and
conducted extensive parameter tuning to find the best parameters for the smart charging use case.

The paper is structured as follows.
First, we explain each aggregation/disaggregation method in detail (\Cref{sec:methods}).
Then, we introduce the benchmarking setup (\Cref{sec:setup}).
Finally, we conduct parameter tuning and determine the performance of each method (\Cref{sec:results}).
In \Cref{sec:conclusion}, we draw conclusions from the results.

The benchmarking setup and our implementations of the different aggregation-disaggregation methods are available on Github: \url{https://github.com/L-Strobel/bench_ev_agg_disagg}

\section{Methodologies}
\label{sec:methods}

This section describes the aggregation/disaggregation methods for \ac{EV} charging events that we evaluate in this paper.
We focus on the most commonly applied methods (\ac{REP} and \ac{VB}).
Many methods described in the literature are variations of the \ac{VB} concept
\cite{pertlEquivalentTimeVariantStorage2019,zhouFormingDispatchableRegion2021,gunkelPassiveActiveFlexibility2020,vayaSmartChargingPlugin2012,haoCharacterizingFlexibilityAggregation2014,abgottsponScalingManagingLarge2018},
their differences will be discussed in \Cref{sec:vb}.
Additionally, we include the \ac{FO} and \ac{DFO} methods because they address specific shortcomings of the common approaches
(see \Cref{sec:FO} and \Cref{sec:DFO}).

In \Cref{sec:otherMethods},
we will briefly overview other methods in the literature that have not yet been included in the benchmark
and justify their exclusion.

\subsection{Representative Profile}
\label{sec:rp}
The \acf{REP} method models the entire fleet by picking a subset of the vehicles and scaling their parameters
(battery capacity, rated charging power, consumption, etc.) so that the sum of the profiles is the same as that of the original fleet in each parameter category.
This approach is very straightforward and has been used in numerous energy system studies \cite{schlundElectricVehicleCharging2021,strobelJointAnalysisRegional2022,schillPowerSystemImpacts2015}.

It can be argued whether the \ac{REP} approach is an aggregation method.
On the one hand, the approach describes the fleet with a reduced set of variables.
On the other hand,
it does not compress the information of the entire fleet, and the resulting estimated load is usually not meant to be disaggregated back to the entire fleet (although that is possible).
Nonetheless, the \ac{REP} approach does represent an important baseline against which the other approaches can be evaluated.

\subsection{Virtual Battery}
\label{sec:vb}

The \ac{VB} method models the \ac{EV} fleet as a single energy storage.
In the literature, many different specifications have been presented \cite{pertlEquivalentTimeVariantStorage2019,zhouFormingDispatchableRegion2021,gunkelPassiveActiveFlexibility2020,vayaSmartChargingPlugin2012,brinkelSchedulingElectricVehicle2022}.
In all cases, the representative energy storage is defined by the time-dependent maximum and minimum power and energy content,
as well as the energy consumption resulting from the mobility behavior.
The implementations differ in their representation of the current energy state of the \ac{VB}.
These differences have minor performance implications.
Therefore, we will use the most popular approach of summing the energy contents of all constituents in each time step.

\begin{figure}
    \centering
    \includegraphics[width=0.9\columnwidth]{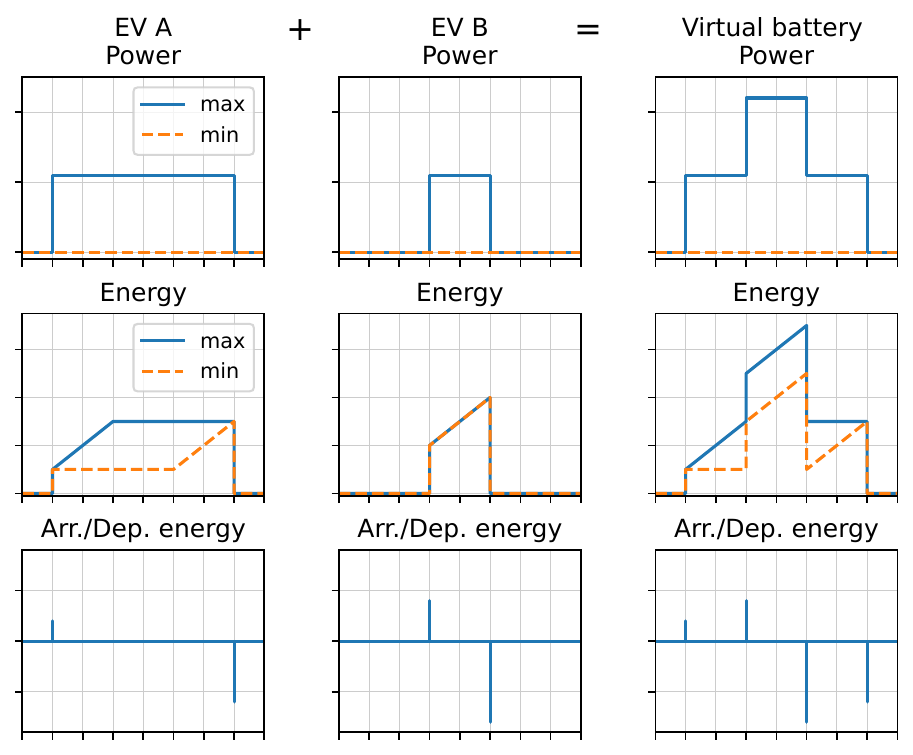}
    \caption{Aggregation of two \ac{EV} charging events into a \ac{VB}.
    First row, determine the power bounds of the \ac{VB}.
    Second row, determine the energy bounds of the \ac{VB}.
    Third row, determine arriving and departing energy.}
    \label{fig:vb}
\end{figure}

The \ac{VB} is defined in three steps, visualized in \Cref{fig:vb}.
First, the minimum and maximum power of the battery is modeled by adding the minimum and maximum charging power of each plugged-in \ac{EV}.
Second, each plugged-in \ac{EV}'s minimum and maximum energy content is determined and aggregated.
The maximum energy content results from charging the vehicles as fast as possible.
The minimum energy content results from charging the vehicles as late as possible.
Lastly, the energy content of arriving \acp{EV} is modeled with a time series that states at which time steps and with what energy content \acp{EV} plug in.
Conversely, a departing energy time series is created from the energy contents when they plug out.

Once the \ac{VB} is defined, it is included in an optimization model with the following constraints:

\begin{equation}
    e_{t}^{VB, min} \leq e_{t}^{VB} \leq e_{t}^{VB, max} \quad \forall t \in T
    \label{eq:cnstr1}
\end{equation}

\begin{equation}
    p_{t}^{VB, min} \leq p_{t}^{VB} \leq p_{t}^{VB, max} \quad \forall t \in T
    \label{eq:cnstr2}
\end{equation}

\begin{equation}
    e_{t+1}^{VB} = e_{t}^{VB} + p_{t}^{VB} \cdot \Delta t \cdot \eta + arr_{t} - dep_{t} \quad \forall t \in T
    \label{eq:cnstr3}
\end{equation}

Where $e_{t}^{VB}$ is the energy content of the \ac{VB} at time t,
$p_{t}^{VB}$ is the power draw of the \ac{VB},
$\eta$ is the charging efficiency,
and $T$ is the set of time steps.
The values $arr_{t}$ and $dep_{t}$ represent the cumulative energy content of \acp{EV} that plug in or out, respectively.

The electric load resulting from this optimization can be disaggregated onto each individual \ac{EV} with a second optimization or, more commonly, with a heuristic method.
Different heuristic methods are proposed in the literature
\cite{schlundFlexAbilityModelingMaximizing2020,nakahiraSmoothedLeastlaxityfirstAlgorithm2017,liRealtimeFlexibilityFeedback2020a,haoCharacterizingFlexibilityAggregation2014,subramanianRealtimeSchedulingDeferrable2012,liuSchedulingAlgorithmsMultiprogramming1973}.
The most commonly used are the least-laxity-first \cite{schlundFlexAbilityModelingMaximizing2020,subramanianRealtimeSchedulingDeferrable2012} and earliest-departure-first algorithms \cite{liuSchedulingAlgorithmsMultiprogramming1973}.
Both methods sort the \acp{EV} and then charge them at full power one after another until the desired electric load is reached.
The difference between the algorithms is how the \acp{EV} are sorted.
In the least-laxity-first algorithm, the \acp{EV} are sorted based on how long they can delay charging while still reaching their \ac{SOC} target at the end of the charging event.
The vehicles that can wait the least time will be charged first.
In the earliest-departure-first algorithm, the \acp{EV} will be sorted according to their departure time.

Regardless of the disaggregation algorithm,
all standard \ac{VB} formulations overestimate the flexibility of the \acp{EV} (see \Cref{apdx:proof} for a proof).
Because of this, there is no guarantee that the \ac{EV} fleet can reproduce the optimized electric load.

Hao and Chen \cite{haoCharacterizingFlexibilityAggregation2014} created an exact \ac{VB} formulation
where this flaw is resolved and perfect disaggregation is possible.
However, their formulation only applies when all \acp{EV} arrive and depart at exactly the same time and have the same rated power.

Abgottspon et al. \cite{abgottsponScalingManagingLarge2018} propose the inclusion of another constraint into the \ac{VB} formulation
that limits the maximum power of the \ac{VB} based on its \ac{SOC}.
The constraint is defined with a function $f(SOC)$ that takes the current \ac{SOC} as input and outputs the percentage of the \acp{EV} aggregated charging power that the \ac{VB} can use (see \Cref{eq:cnstrFPC}).

\begin{equation}
    \frac{p_{t}^{VB}}{p_{t}^{VB, max}} \leq f(SOC_t) \quad \forall t \in T
    \label{eq:cnstrFPC}
\end{equation}

where

\begin{equation}
    SOC_t = e^{VB}_t / e^{VB,max}_t
    \label{eq:soc}
\end{equation}

\begin{figure}
    \centering
    \includegraphics[width=0.9\columnwidth]{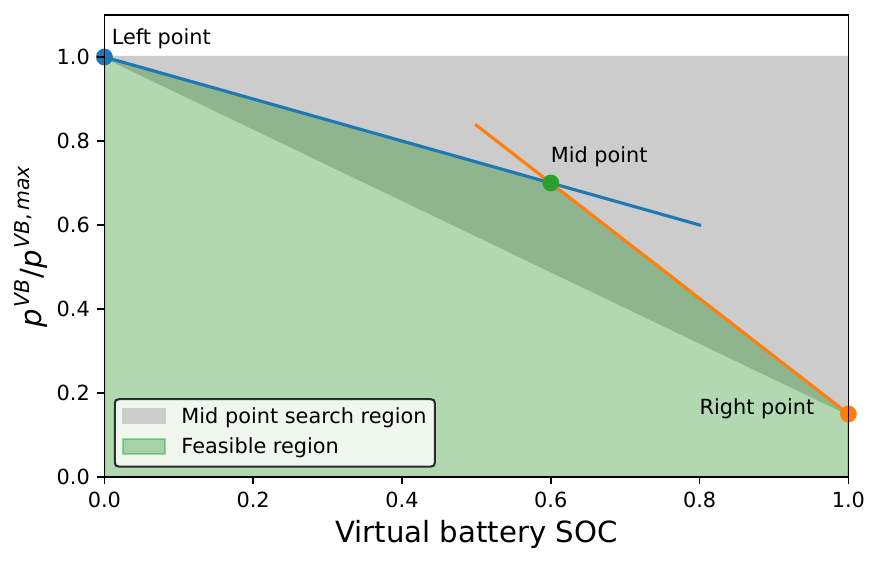}
    \caption{Optional additional piece-wise linear constraint for the \ac{VB}.
    For illustration purposes, the right point is depicted as already determined.}
    \label{fig:fpcexp}
\end{figure}

We can use this constraint
to limit the overestimation of the \ac{VB}
by reducing the charging power at high \ac{SOC},
where some of the constituent \acp{EV} might be fully charged and can no longer contribute their
power.

In this work, we will choose a piece-wise linear formulation of $f(SOC)$ as depicted in \Cref{fig:fpcexp}.
$f(SOC)$ is defined by three points.
The left point determines the maximum power if the \ac{VB} is empty.
If the \ac{VB} is empty,
it can be charged with all the available power, so the left point is always (0, 1).
The middle point determines the intermediate behavior of $f(SOC)$.
We determine its location through the parameter tuning process described in \Cref{sec:pm}.
To keep the optimization linear, we require the middle point to be above the line between the left and right points.
The right point determines the maximum power with which the last bit of energy that fills the \ac{VB} can be charged.
The x-value of the right point is always 1; its y-value is determined through parameter tuning.

In this work,
we benchmark the performance of the following \ac{VB} formulations:

\begin{itemize}
    \item VB-LL: Standard formulation of \ac{VB}. Disaggregation with the least-laxity-first algorithm.
    \item VB-ED: Standard formulation of \ac{VB}. Disaggregation with the earliest-departure-first algorithm.
    \item VB-LL-FPC: Standard formulation of \ac{VB} with additional piece-wise linear constraint \Cref{eq:cnstrFPC} (see. \cite{abgottsponScalingManagingLarge2018}).
    Disaggregation with the least-laxity-first algorithm. 
    \item VB-LL-Grpd: Multiple \acp{VB}. Charging events are first grouped with the \ac{FO} grouping method proposed in \cite{siksnysAggregatingDisaggregatingFlexibility2015a},
    then each group is aggregated to a single \ac{VB}. The standard formulation is used for each \ac{VB}. Disaggregation with the least-laxity-first algorithm.
\end{itemize}

\subsection{Flex Objects}
\label{sec:FO}
The \ac{FO} concept \cite{siksnysAggregatingDisaggregatingFlexibility2015a}
(also known as standard \ac{FO} to differentiate from later extended versions \cite{pedersenModelingManagingEnergy2018a,lilliuUncertainFlexOffersScalable2023})
is an abstract representation of flexible energy prosumers
that enables us to model multiple such prosumers, with possibly very different characteristics, in a unified format capable of aggregation and disaggregation.

The approach is geared towards real-world usage and, contrary to the \ac{VB} approach, allows for error-free disaggregation.
Meaning that it is guaranteed that any load time series feasible for the aggregated object can be disaggregated exactly onto its constituents.
This property is highly desirable in a real-world context
because it ensures that all market offers made by an aggregator can be fulfilled.
Similarly, the property is also desirable for a unit commitment model
because it guarantees that any optimization solution of an aggregated model is feasible for the original model.
However, this advantage comes at the cost that the flexibility is generally underestimated.
To what degree, depends on the number of aggregated objects. 
Increasing the number of aggregated objects will reduce the underestimation at the cost of additional runtime.

\begin{figure}
    \centering
    \includegraphics[width=0.9\columnwidth]{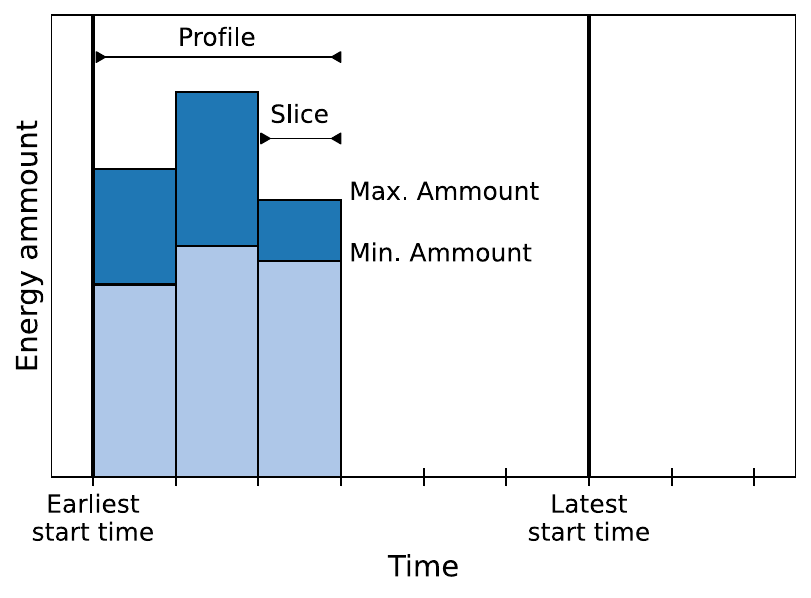}
    \caption{Visualization of a Flex Object. This figure is based on a visualization in \cite{siksnysAggregatingDisaggregatingFlexibility2015a}.}
    \label{fig:fo}
\end{figure}

A \ac{FO} describes the flexibility of a prosumer in terms of a profile defined by the maximum and minimum energy
they can generate/consume in discrete consecutive time intervals called slices (see \Cref{fig:fo}).
The profile's start time is flexible within the range of the earliest start time $t_{es}$ and the latest start time $t_{ls}$.
Therefore, a \ac{FO} is entirely defined by the tuple $(t_{es}, t_{ls}, p)$
where $p$ is the profile defined as a sequence of slices: $p=\langle s_1, ... , s_n \rangle$.
Each slice $s$ is a tuple that specifies the maximum and minimum energy amount: $s = (e_{min}, e_{max})$.

With this definition, the \ac{FO} has two sources of flexibility:
Firstly, the energy flexibility of each slice and, secondly, the time flexibility defined by $t_{ls} - t_{es}$.
We utilize these flexibilities by setting the \ac{FO}'s start time and each slice's energy within the given limits.
This process is called instantiation \cite{siksnysAggregatingDisaggregatingFlexibility2015a}.

Aggregated \acp{FO} are built from a set of individual \acp{FO} via an algorithm explained in detail in \cite{siksnysAggregatingDisaggregatingFlexibility2015a}.
At its core, the aggregation algorithm fixes the start times of each constituent \ac{FO} relative to the start of the aggregated \ac{FO}.
For example, the start time of a constituent might be set to $t + 2$, where $t$ is the start time of the aggregated \ac{FO}.
This way, what slices occur simultaneously is certain, and those that coincide can be aggregated by adding their minimum and maximum amounts.
The aggregated \ac{FO} is treated in the same manner as a basic \ac{FO}.

This aggregation method loses a significant amount of the \ac{FO}'s flexibility
because it sets the time flexibility of the aggregated \ac{FO} to that of its least flexible constituent.
Therefore, it is necessary to aggregate the \acp{FO} into more than one aggregated \ac{FO},
each consisting of similar \acp{FO}.
Siksnys et al. \cite{siksnysAggregatingDisaggregatingFlexibility2015a} do this by grouping the \ac{FO}
with an algorithm that guarantees certain attributes (like earliest start time, latest start time, or time flexibility)
of the \ac{FO} differ at most by a specified threshold for each group.
The details of this grouping mechanism can be found in \cite{siksnysAggregatingDisaggregatingFlexibility2015a}.

Once the aggregated \ac{FO} is instantiated, it can be disaggregated.
The start time of each constituent \ac{FO} is set according to its relative position in the aggregate.
The value of each slice is set to the same relative position within its bounds as that of the aggregated slice.
For example, if a slice of the aggregated \ac{FO} $s^a=(0,10)$ is instantiated with the value 2,
then a constituent slice $s^i=(0,5)$ will be instantiated with 1
so that both are at 20 \% usage.

We implemented the \ac{FO} approach as follows.
Each \ac{FO} represents one charging event.
Since each charging event has a fixed \ac{SOC} upon departure,
it must be that $e_{min} = e_{max}$ for all slices.
This is necessary because, in the form presented in \cite{siksnysAggregatingDisaggregatingFlexibility2015a},
Standard \acp{FO} do not include a cumulative energy constraint that specifies the energy that must be delivered
over multiple time slices.
Therefore, to ensure that the necessary energy is charged in each event it must be that $e_{min} = e_{max}$.
Otherwise, an optimization could  always charge at $e_{min}$ and thereby not reach the desired total energy.
We construct the values for $e_{min}/e_{max}$ via a simulation of uncontrolled charging (i.e., charging as fast as possible after plug-in. See \Cref{sec:setup}).
The resulting energy demand in each time slice defines the profile of the \ac{FO}.
The earliest start time is the start time of uncontrolled charging,
while the latest start time is computed from the charging duration and the idle time after charging.
For grouping, we use thresholds on the earliest start time and time flexibility (as in the experimental evaluation of \cite{siksnysAggregatingDisaggregatingFlexibility2015a}).
\footnote{
    We skip the bin-packing phase of the grouping algorithm
    because its purpose of balancing the aggregated  \acp{FO} is irrelevant to our benchmark.
}

\subsection{Dependency-based FlexOffers}
\label{sec:DFO}

\acp{DFO} \cite{siksnysDependencybasedFlexOffersScalable2016} are a derivative of the \ac{FO} model.
They differ from \acp{FO} by making each slice's minimum and maximum energy amounts dependent on the total energy consumed in previous slices.
This means that instead of a slice being represented by a 1-dimensional range;
each slice is defined by a polygon where the y-axis indicates how much energy can be consumed during the time step
and the x-axis states how much energy was consumed previously.
The area inside the polygon defines the feasible region of the slice.

The additional dependency on previous slices
allows \acp{DFO} to model multiple real-world restrictions that could not be modeled with \acp{FO}.
For example, it is possible to model the maximum allowed deviation from a target temperature in a heat pump flexibility model.
Similarly, we can model a fixed total cumulative energy consumption for a charging event.

\begin{figure}
    \centering
    \includegraphics[width=0.9\columnwidth]{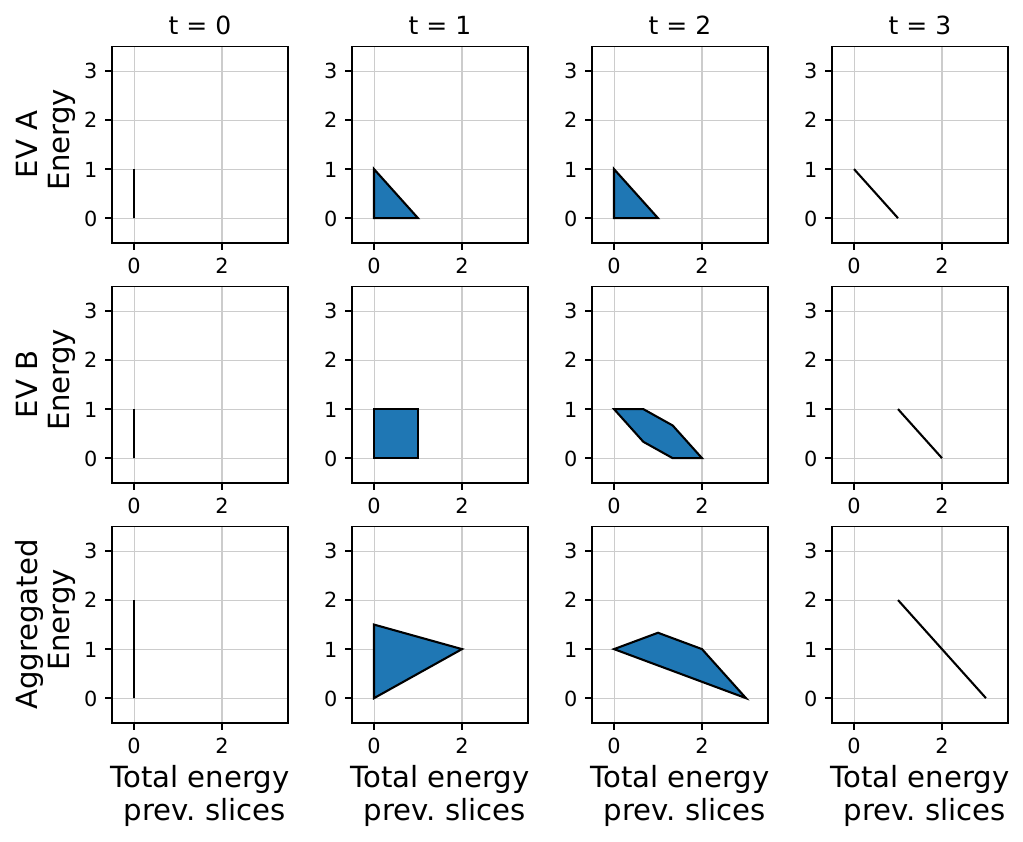}
    \caption{\ac{DFO} aggregation. The first two rows are \acp{DFO} that encode the flexibility of two individual charging events.
    The last row shows the aggregation of the two.
    The figure is based on a visualization in \cite{siksnysDependencybasedFlexOffersScalable2016}.}
    \label{fig:dfo}
\end{figure}

The first row of \Cref{fig:dfo} depicts an example of a \ac{DFO} for a charging event.
The \ac{EV} is plugged in in the first time step (t = 0); since no charging was possible beforehand, the x-values of all polygon points must be 0.
The maximum charging rate and the time step length define the y-value’s range
At t=1 and t=2, the energy charged previously and during the time step is variable.
Therefore, the feasible region is defined by a polygon.
In our benchmark, the total energy of a charging event is fixed.
Consequently, the energy that can be charged in the last step entirely depends on the previous steps, reducing the feasible region to a line.

The aggregation process of \acp{DFO} is more complicated than that of \acp{FO}.
We refer to the original paper \cite{siksnysDependencybasedFlexOffersScalable2016} for a detailed explanation.
Generally, we always aggregate two \acp{DFO}, A and B, at once.
Larger aggregations are achieved by aggregating the result with another \ac{DFO} C and so on.
The basic concept is to iterate through the slices of A and B simultaneously.
For each pair of slices, the minimum and maximum y values are evaluated at $n$ equally spaced points in the range $[0, 1]$
that represent relative positions within the x-axis bounds of the slice.
At each relative position, the minimum and maximum y values of the aggregate \ac{DFO} are obtained by summing the minimum and maximum y values of both slices.
The absolute x-bounds of the aggregate are defined by the highest and lowest $l_1$-norms over all feasible points of the previous slice
(The x-bounds of the first slice are $[0,0]$).

The value $n$ represents the accuracy with which we determine the aggregated \ac{DFO} and will be called $NumSamples$ from here on.

Like \acp{FO},
the \ac{DFO} disaggregation process determines the relative usage of the total energy bounds of the aggregated \ac{DFO} at each time step
and sets the relative usage of the constituents to the same level.
In certain cases, this is not possible with the aggregation method described above.
In these cases, the \ac{DFO} aggregation algorithm reduces the y-range of the aggregate where necessary.
This ensures error-free disaggregation at the cost of flexibility.
See \cite{siksnysDependencybasedFlexOffersScalable2016} for more details.

Like with \acp{FO}, the flexibility loss depends on the similarity of the underlying prosumers.
Therefore, it is again necessary to aggregate to multiple aggregated \acp{DFO} that consist of similar \acp{DFO}.
To find similar \acp{DFO}, we utilize the same grouping algorithm as for \acp{FO} \cite{siksnysAggregatingDisaggregatingFlexibility2015a}.\footnote{
    Siksnys et al. \cite{siksnysDependencybasedFlexOffersScalable2016} propose a grouping method
    that measures the similarity of \acp{DFO} with the bound reduction necessary in step 5 of the aggregation algorithm.
    However, this method yielded a similarity of 0 \% for most charging events we analyzed.
    Because of this, we could not reach the desired compression,
    leading us to use the \ac{FO} grouping method instead.
}

Our implementation of the \ac{DFO} creation process from charging events is straightforward;
the energy bounds follow directly from the limits on the charging power and the fixed cumulative energy amount required for the entire charging event.

For benchmarking, we integrate the aggregated \acp{DFO} into a linear optimization using the geometric viewpoint.
At each slice, we determine a set of lines that define the boundary of that polygon.
Since \ac{DFO} slices are always convex, we can add one constraint for each line to the optimization that ensures that a point
is on the correct side of the line and, therefore, inside the polygon.
Each point is represented by two variables:
one for its x-value (the charging amount at a given time)
and one for its y-value (the total charging amount up to that time).
For each point, we introduce another constraint that enforces equality between its y-value and the x-value of all  previous points x-values.
The complexity of this optimization depends mainly on the number of lines needed to describe each slice,
which in turn strongly depends on the $NumSampels$ parameter.

\subsection{Other Methods}
\label{sec:otherMethods}

Apart from the methods included in the benchmark,
there are several other methods we found in the literature.

Multiple studies take the viewpoint of an aggregator of deferrable loads participating in flexibility markets
\cite{liLearningBasedPredictiveControl2021a,tsaousoglouFlexibilityAggregationTemporally2022}.
These studies aim to find aggregated representations of the flexibility of their portfolios
to communicate their flexibility to the system operator and to allow for effective bidding on flexibility markets.
Often, the focus is on signals passed from the system operator to the aggregator and from the aggregator to individual flexible devices.
Li et al. \cite{liLearningBasedPredictiveControl2021a} approach this problem by introducing a \textit{penalized predictive control}
algorithm that utilizes a \textit{maximum entropy feedback}.
Tsaousoglou et al. \cite{tsaousoglouFlexibilityAggregationTemporally2022} train machine learning algorithms that
receive information about the flexible resources,
submit biddings to the system operator,
and disaggregate the feedback of the system operator back onto the flexible resources.
Both approaches are highly specialized towards real-time online control of flexible resources,
and it would be necessary to take significant liberties in their implementation to include them in our offline optimization benchmark.
For this reason, they are not included.

The FlexAbility approach proposed by Schlund et al. \cite{schlundFlexAbilityModelingMaximizing2020}
is another online aggregation method.
The method aggregates the flexibility of an \ac{EV} fleet by computing how long the fleet can supply a certain set of powers $P$.
A set of tuples $(power, duration)$ then describes the aggregated flexibility of the fleet at time $t$.
At time $t+1$ this set has to be recalculated while taking the realized power at $t$ into account.
The online nature of the method again limits its comparability with the methods we benchmark in this paper.

Next to the \ac{DFO} method,
several other methods use the Polytop viewpoint of flexibility
\cite{ulbigAnalyzingOperationalFlexibility2015,trangbaekExactConstraintAggregation2012,barotConciseApproximateRepresentation2017,zhaoExtractingFlexibilityHeterogeneous2016,mullerAggregationDisaggregationEnergetic2019,mullerAggregationEnergeticFlexibility2015}.
Generally, the aggregation of a set of polytopes representing flexibilities is computed via its Minkowski sum.
However, computing the Minkowski sum of general polytopes is computationally demanding \cite{mullerAggregationDisaggregationEnergetic2019,zhaoExtractingFlexibilityHeterogeneous2016,tiwaryHardnessComputingIntersection2008}.
Therefore, most of these methods use either an approximation of the Minkowski sum 
that can be computed efficiently \cite{zhaoExtractingFlexibilityHeterogeneous2016,barotConciseApproximateRepresentation2017}
or use special polytopes \cite{mullerAggregationDisaggregationEnergetic2019,mullerAggregationEnergeticFlexibility2015}.
For example, Müller et al. propose a method of approximating flexibility objects as Sonotopes \cite{mullerAggregationEnergeticFlexibility2015}.
This formulation allows for a simple calculation of the Minkowski sum \cite{mullerAggregationDisaggregationEnergetic2019}.
For compactness, we decided only to include one polytope  method in the first version of our benchmark.
The \ac{DFO} method was the natural choice because it is an extension of the popular \ac{FO} approach.
In future work, we also aim to benchmark the other polytope methods.

\section{Benchmark Setup}
\label{sec:setup}

The benchmark setup evaluates how well the presented methods in \Cref{sec:methods} reproduce the true benefit of smart charging,
compared to uncontrolled charging.
The key metric here is the charging costs.
More precisely, the \textit{gap} between the charging cost in the true optimal case (OPTIMAL),
where each vehicle is optimized individually (see \Cref{apdx:opt}),
and the case where one of the methods is used to aggregate/disaggregate the charging events before/after optimization.

This \textit{gap} is calculated as follows:

\begin{equation}
    gap = \frac{|costs_{Method} - costs_{Optimal}|}{|costs_{Optimal}|}
    \label{eq:gap}
\end{equation}

The charging costs are defined as follows:

\begin{equation}
    costs = \sum_{t \in T}{\sum_{v \in V}{p_{t, v} \cdot \Delta t \cdot c_{t} }}
    \label{eq:costs}
\end{equation}

Where $T$ is the set of time steps in the time horizon (one week in all tested cases),
and $V$ is the set of \acp{EV}.
$c_t$ is the cost of charging at time step $t$,
$p_{t, v}$ is the power drawn by vehicle $v$ at time step $t$,
$\Delta t$ is the time step length.

\begin{figure}
    \centering
    \includegraphics[width=0.9\columnwidth]{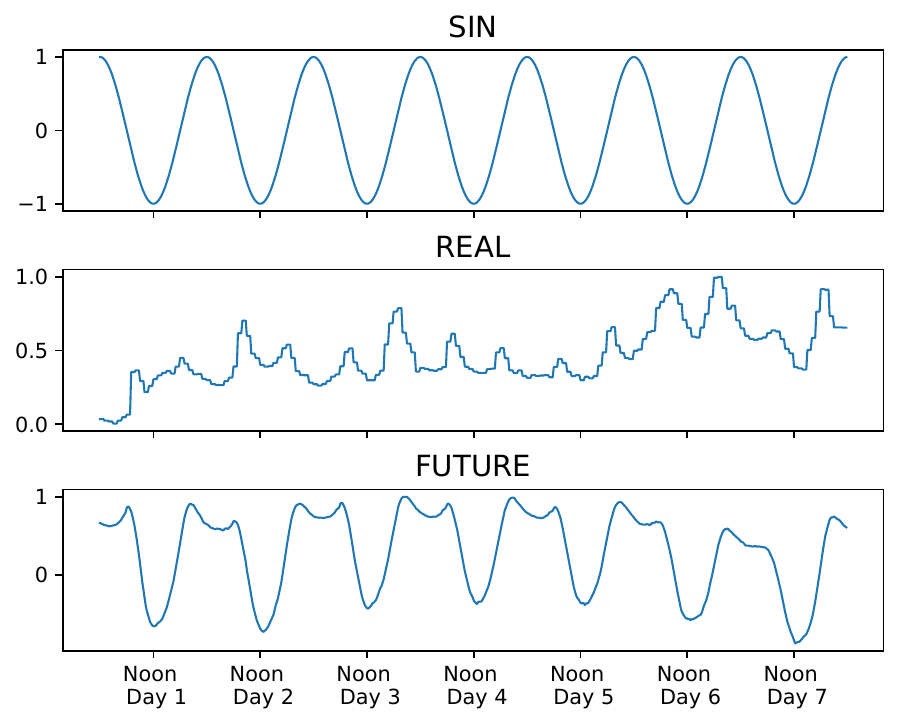}
    \caption{Examples for the price time series}
    \label{fig:ps}
\end{figure}

We use the three price time series for $c_t$ depicted in \Cref{fig:ps}.
All of them represent electricity prices:

\begin{itemize}
    \item SIN:
    A sine wave with a period of one day and an amplitude of one.
    We chose it because it is the simplest way to model the daily price fluctuations typical for electricity spot market.
    It is helpful to see how each method performs for the simplest non-trivial case.
    \item REAL:
    The real spot market price of a random week in Germany in 2022 \cite{bundesnetzagenturSMARDStrommarktdaten2023}.
    The price depicted in \Cref{fig:ps} is only an example;
    a new random week is selected for each repetition.
    \item FUTURE:
    An estimation of the residual load (el. demand - generation of fluctuating renewables) in Germany in 2045.
    The price time series is created by scaling the demand, PV generation, and wind generation of 2022 with the total demand and installed capacity
    predicted for 2045 (in Scenario B of the German grid development plan 2037/2045 (2023) \cite{50hertztransmissiongmbhNetzentwicklungsplan203720452023}).
    We choose a new random week for every repetition.
    This time series represents a rough estimation of price fluctuations in future energy systems.
    The motivation behind this choice is that electricity is likely cheap if renewable generation is abundant and expensive if the demand is high\footnote{
    The residual load already is a factor behind electricity prices.
    In 2022, the correlation between residual load and electricity prices in Germany was 0.61 \cite{bundesnetzagenturSMARDStrommarktdaten2023}.}.
\end{itemize}

Each price signal is normalized so that its maximum value is one.

The driving behavior of the \acp{EV} is modeled with driving profiles
from the data set \ac{MID} \cite{infasMobilitatDeutschland20172017}.
The \ac{MID} is Germany's large-scale household travel survey.
The \ac{MID} was created on behalf of the German Federal Ministry for Digital and Transport and is one of the largest surveys about everyday mobility.
Its purpose is to supply representative and reliable mobility data that serves as the basis for local and national transportation models
and as the foundation for policy decisions \cite{infasMobilitatDeutschland20172017}.
The data set includes detailed information (like distance, departure time, arrival time, and transport mode)
about all trips a participant undertook on one randomly chosen day in 2017.

We create daily driving profiles for each \ac{EV} by selecting a random individual from the data
and building a chain from their car trips.
In between trips, the \ac{EV} parks.
The location of parking is dependent on the purpose of the previous trip.
For example, if the purpose of the previous trip was "to work",
then the vehicle will be considered "At work" until the next trip.
Driving profiles for longer periods are created by combining multiple such driving profiles
with the condition that the weekdays of the profiles are in order.
We first sample a profile for a Monday, then a Tuesday, and so on.
Through this process, we obtain weeklong driving profiles represented by a list of consecutive driving and parking events that together contain all the necessary information,
i.e., vehicle locations, trip and park durations, and distances traveled.

The advantage of this commonly used method for generating driving profiles is that the profiles
are based on real data representative of the entire population of a country,
and their generation requires minimal assumptions.
Multiple open-source tools used to model \ac{EV} loads in the energy system modeling field
utilize similar  household travel survey-based approaches \cite{gaete-moralesOpenToolCreating2021a,wulffVehicleEnergyConsumption2021,strobelOMODOpensourceTool2023}.
The disadvantage is that due to a lack of data,
most profiles are created based on the driving profiles of internal combustion engine vehicles.
However, as demonstrated in \cite{strobelJointAnalysisRegional2022},
the vast majority of everyday car trips are compatible with the ranges of ordinary \acp{EV}.
Therefore, in the absence of necessity, it is unlikely that everyday driving behavior will change.
Long-distance driving behavior is another matter.
Here, the \ac{MID} is not necessarily representative.
Consequently, our results are less meaningful for highway charging,
where flexibilities are likely far smaller than in typical everyday mobility patterns.
Another disadvantage of generating driving profiles from the \ac{MID} is that the behavior of each \ac{EV} is based on a different person every day since,
for every participant,
only one day of data exists.

\begin{table}
    \centering
    \begin{tabular}{c c}
        Name & Value \\
        \hline
        $\Delta t$               & 15 min \\
        \ac{SOC} at start        & 0.9 \\
        \ac{SOC} at end          & 0.9 \\
        Charging effiency        & 90 \% \\
        Charging location        & Only at home \\
        Maximum charging power   & 11 kW \\
        Battery capacity         & 72 kWh \\
        Consumption              & 14.5 kWh/100km \\
    \end{tabular}
    \caption{\ac{EV} parameters}
    \label{tab:Param}
\end{table}

The \ac{EV} parameters are summarized in \Cref{tab:Param}.
They are based on the medium-sized \ac{EV} in \cite{strobelJointAnalysisRegional2022}
to represent average vehicles in a future setting with large numbers of \acp{EV}.
Charging will only be allowed at home to simplify assumptions for charging point availability and maximum charging power.

To allow for a fair comparison between methods,
we introduce certain constraints on the flexibility of \ac{EV} charging.
\begin{enumerate}
    \item For any given charging event, an \ac{EV} must charge the same energy amount as in the uncontrolled case.
    Therefore, no energy shift between locations is possible.
    \item \Ac{V2G} charging is not allowed.
    \item The \ac{SOC} at the end of the time horizon must be the same as at the beginning.
    If that is not possible, it must be the same as in the uncontrolled case.
\end{enumerate}

Constraint (1) lets us view each charging event as an independent flexibility event.
Otherwise,  energy could be shifted over the entire time horizon
and the time horizon would be one single flexibility event.
This raises serious questions about how to implement the different aggregation methods.
For example, with \acp{FO}, the parameter "earliest start time" ($t_{es}$) is meaningful for one charging event,
but not for a weeklong driving profile.
\acp{DFO} can be formulated without constraint (1),
but the method performs badly if time slices with no flexibility (e.g. \ac{EV} is driving)
and slices with flexibility are aggregated into one \ac{DFO}.
Constraint (2) is introduced because of the missing cumulative energy constraint for \acp{FO}.
Constraint (3) has nothing to do with the shortcomings of any algorithm
but simply ensures that no method is rewarded for charging less energy than another.

To put the performance of each method into perspective,
we also evaluate the \textit{gap} of uncontrolled charging (UNCONTROLLED),
where \acp{EV} always charge with the maximum charging rate as soon as they arrive at a location with an available charging point.

\section{Results} 
\label{sec:results}

\subsection{Parameter tuning}
\label{sec:pm}

\begin{figure*}
    \centering
    \begin{subfigure}[c][][c]{1\textwidth}
        \centering
        \includegraphics[width=1\columnwidth]{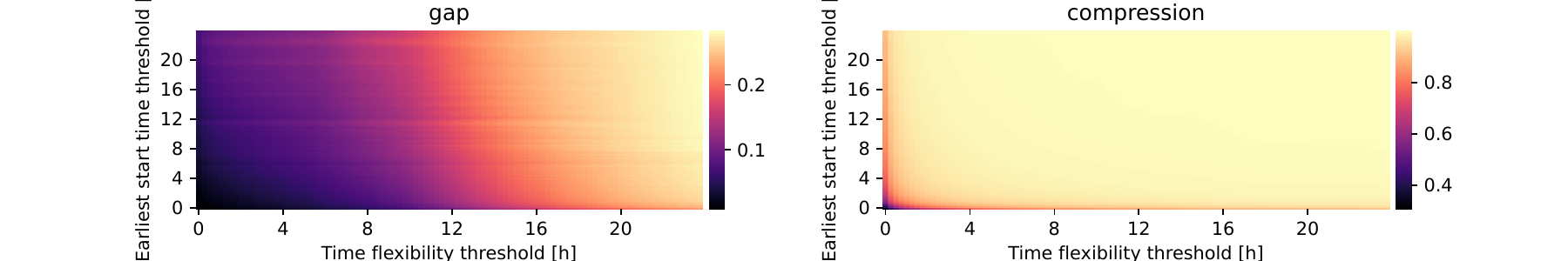}
        \caption{\acf{FO}}
        \label{fig:fopm}
    \end{subfigure}
    \begin{subfigure}{1\textwidth}
        \centering
        \includegraphics[width=1\columnwidth]{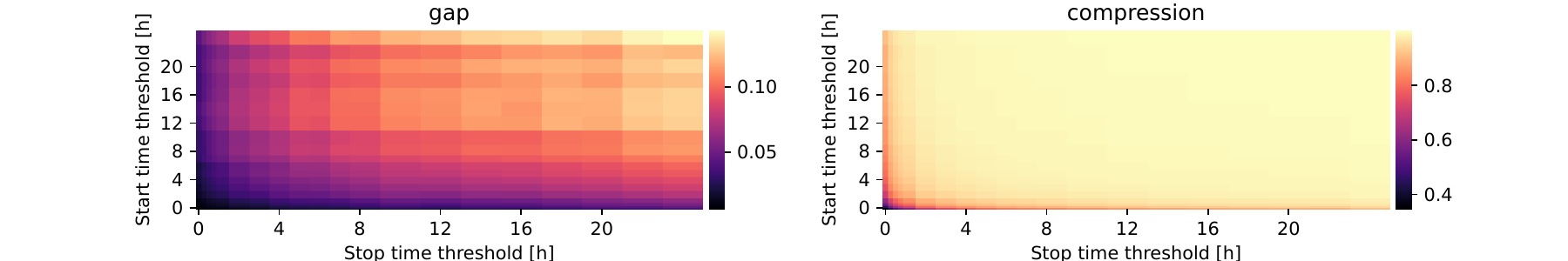}
        \caption{\acf{DFO}}
        \label{fig:dfopm}
    \end{subfigure}
    \begin{subfigure}{1\textwidth}
        \centering
        \includegraphics[width=1\columnwidth]{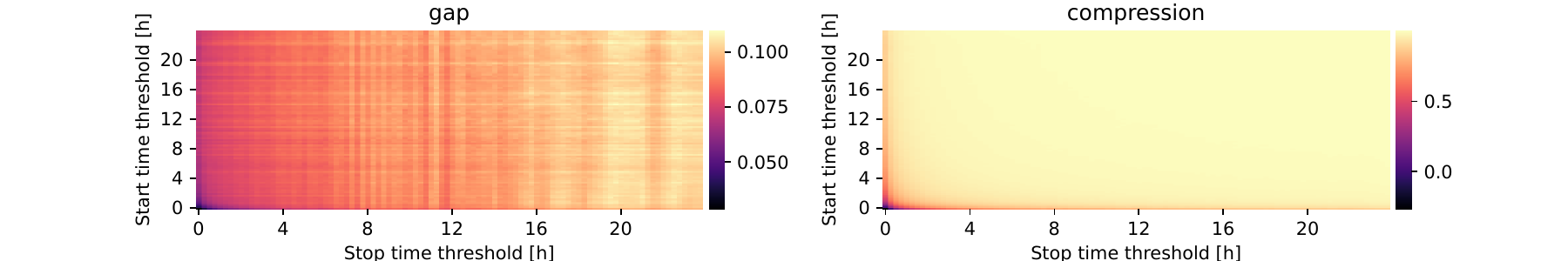}
        \caption{Virtual battery grouped (VB-LL-Grpd)}
        \label{fig:vbgrpdpm}
    \end{subfigure}
    \begin{subfigure}{1\textwidth}
        \centering
        \includegraphics[width=1\columnwidth]{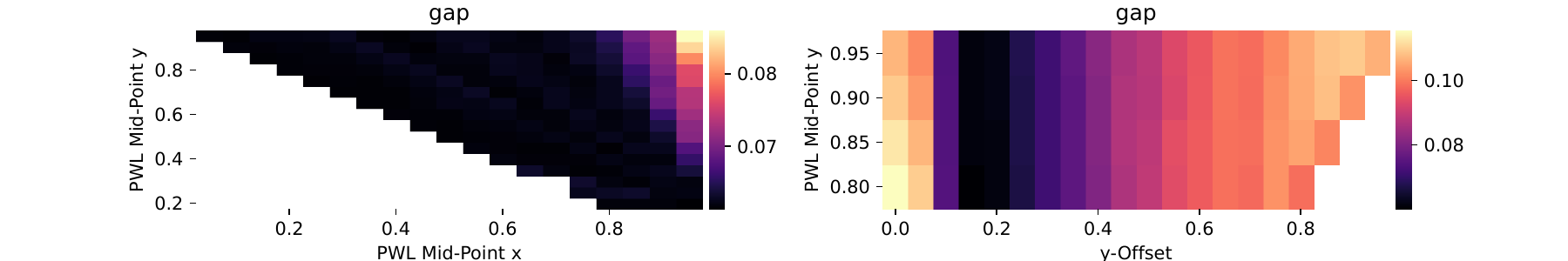}
        \caption{Virtual battery with piece-wise linear constraint (VB-LL-FPC)}
        \label{fig:vbfosepm}
    \end{subfigure}
    \label{fig:pm}
    \caption{Parameter tuning}
\end{figure*}

Most of the discussed algorithms have parameters that need to be tuned.
In this section, we conduct this parameter tuning via grid search.
For each parameter set in the search space we compute the \textit{gap} and \textit{compression} ($compression = \frac{n_{AggregatedObjects}}{n_{ChargingEvents}}$).
Since this task has high computational costs, we will evaluate each parameter configuration only for the price time series REAL and with 1000 \acp{EV}.
Each configuration run is repeated five times to reduce the influence of the randomness involved in agent and price time series creation.
The results for \textit{gap} and \textit{compression} are the mean of these five runs. 

\acp{FO} and \acp{DFO} have a set of parameters that control the trade-off between flexibility loss and compression.
In the case of \ac{FO}, these parameters are thresholds that describe the minimum necessary similarity between two flexibility objects
that can be aggregated into a single \ac{FO}.
We use thresholds for the earliest start time ($t_{es}$) and the time flexibility ($tf$),
as recommended in \cite{siksnysAggregatingDisaggregatingFlexibility2015a}.
This means that all aggregated \acp{FO} consist of a set of charging events where for any two charging events in that set,
the plug-in time difference is below the $t_{es}$ threshold,
and the time flexibility difference is below the $tf$ threshold.
For example,
if the $t_{es}$ threshold is 5 hours,
then the difference between plug-in times of two charging events inside the same \ac{FO} is, at most, 5 hours.
We conduct a grid search in the range [0, 24] hours and 15-minute intervals for both thresholds.
The results are depicted in \Cref{fig:fopm}.
As expected, the \textit{gap} increases continuously with increasing thresholds.
The increase is significantly steeper for the $tf$ threshold than for the $t_{es}$ threshold,
indicating that it is more important to group equally flexible objects
than to group objects with a similar start time.
The compression increases equally rapidly with both thresholds.

As stated in \Cref{sec:DFO}, the grouping method for \acp{DFO} proposed in \cite{siksnysDependencybasedFlexOffersScalable2016}
performed poorly for our use case.
Instead, we utilize the same threshold-based grouping as for \acp{FO}.
\acp{DFO} are very sensitive to the start and end times of flexibility events.
Therefore, we use a threshold for the plug-in and plug-out time.
Again, the grid search is conducted in the same [0, 24] hour range.
However, since the runtime of \acp{DFO} is significantly higher,
we will search in 15-minute intervals only from 0 to 1 hour,
then use 1-hour intervals from 1 hour to 8 hours,
and finally continue with 2-hour intervals until the end of the range.
\acp{DFO} have an additional parameter, $NumSamples$,
that determines the accuracy of the feasible region of each slice.
In the grid search,
this parameter is set to four to avoid multiplying the already large configuration space.
Once the other parameters are configured, we will evaluate the impact of $NumSamples$ in \Cref{sec:comp}.
The results of the \ac{DFO} parameter tuning are depicted in \Cref{fig:dfopm}.
We can see that the \textit{gap} for \acp{DFO} is generally lower for similar compression levels than for \acp{FO}.
Both plug-in and plug-out thresholds affect the results similarly.

For the grouped version of the virtual battery (VB-LL-Grpd),
we also use the grouping method for \acp{FO} with plug-in and plug-out thresholds.
We conduct the grid search in a [0, 24] hour range with 15-minute intervals.
Our results are depicted in \Cref{fig:vbgrpdpm}.
This method's performance is less dependent on the parameterization than \acp{FO} or \acp{DFO}.
\textit{Gaps} range from 10 \% to 4 \%.
The \ac{VB} approach also works when all events are grouped into one aggregated object.
Therefore, the benefit of grouping similar events is less extreme than in the \ac{FO} and \ac{DFO} cases.
Additionally, the results are more noisy, with regions of good performance interrupted by regions of bad performance.
These could correspond to cyclical patterns in the data.

The parameter tuning for VB-LL-FPC does not involve finding a good trade-off between compression and performance.
Our goal here is to find the best-performing definition of $f$ in the piece-wise linear constraint $p_{max} \leq f(SOC)$.
This involves finding the best position for the middle point (Mid-Point x, Mid-Point y)
and the y-value of the right point (y-Offset).
\Cref{fig:fpcexp} shows all accepted values for the middle point.
We find the best configuration via grid search within a range [0, 1] and 0.05 intervals for all three values.
All configurations where the middle point would lie below the line between the left and the right point are skipped (see \Cref{sec:vb}).
The results are depicted in \Cref{fig:vbfosepm}.
We can see that the exact location of the middle point is mostly irrelevant as long as its x-coordinate is below 0.8.
This means that the piece-wise linear constraint's main benefit is to avoid high charging powers when the virtual battery is almost full.
Interestingly, the y-Offset has a clear optimum at 0.15,
resulting from the trade-off between reducing errors at high
\ac{SOC} and utilizing all the available flexibility.

\begin{table}
    \centering
    \begin{tabular}{c c c}
        Algorithm & Parameter & Value \\
        \hline
        REP & Number of profiles & 500 \\
        FO & earliest start time thresh. & 3h 45min \\
        FO & time flexbility thresh. & 1h 45min \\
        DFO & plug-in time thresh. & 45min \\
        DFO & plug-out time thresh. & 12h \\
        VB-LL-Grpd & plug-in time thresh. & 8h 15min \\
        VB-LL-Grpd & plug-out time thresh. & 2h \\
        VB-LL-FPC & Mid-Point x & 0.8 \\
        VB-LL-FPC & Mid-Point y & 0.6 \\
        VB-LL-FPC & y-Offset & 0.15 \\
    \end{tabular}
    \caption{Parameter tuning results}
    \label{tab:tuningRslts}
\end{table}

\begin{figure}
    \centering
    \includegraphics[width=0.9\columnwidth]{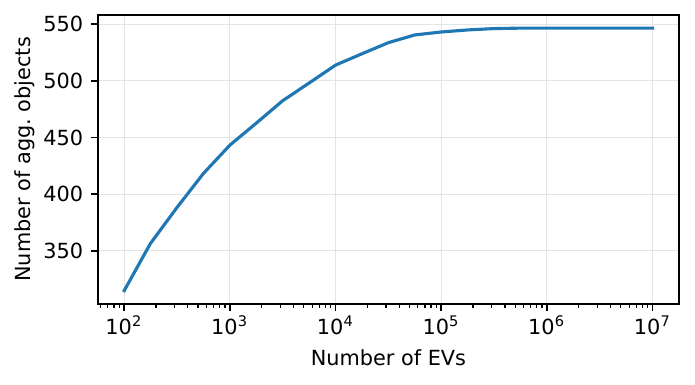}
    \caption{Number of aggregated objects with the \ac{FO} grouping method. Parameterization as for \acp{DFO} (see \Cref{tab:tuningRslts}).}
    \label{fig:compression}
\end{figure}

What is the best set of parameters depends on the available computational resources and time.
Generally, with higher compression,
both the runtime and performance decrease.
However, the marginal improvements will decrease for extreme values of compression.
In this work, we decided to take the best set of parameters,
resulting in a compression of at least 95 \% in the tuning step.
The resulting parameters are summarized in \Cref{tab:tuningRslts}.
This parameterization will result in about 500 aggregated flexibility objects for any number of \acp{EV}
\footnote{This holds for all methods since we always use the \ac{FO} grouping method.};
see \Cref{fig:compression}.
Consequently, we use 500 profiles for the representative profile method (REP).
In VB-LL-FPC, all events are always compressed to one object;
therefore, we simply choose the set of parameters with the lowest \textit{gap}.

\subsection{Method Comparison}
\label{sec:comp}

\begin{figure*}
    \centering
    \includegraphics[width=2.0\columnwidth]{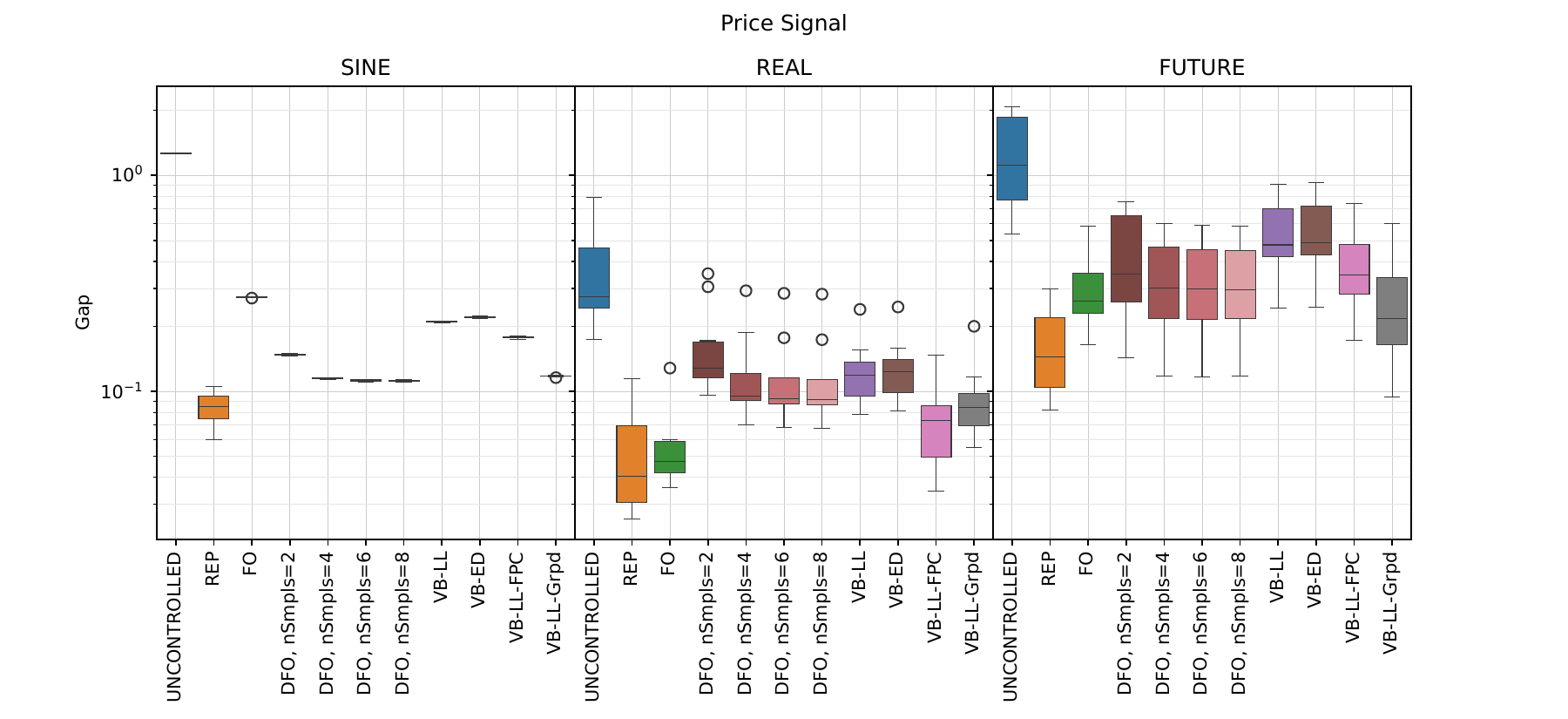}
    \caption{Performance of the different methods on the benchmark for 100,000 \acp{EV}.
    The \textit{gap} is the relative additional cost compared to the optimal case (see \Cref{eq:gap}).}
    \label{fig:rsltGap}
\end{figure*}

The benchmarking results for 100,000 \acp{EV} are depicted in \Cref{fig:rsltGap}.
To compute the gap metric we must compute the real optimal case (OPTIMAL) for each repetition run.
This hinders us from running the benchmark with even higher numbers of simulated \acp{EV}.
However, the results for 100,000 \acp{EV} show only a marginal difference compared to those with 10,000.
Therefore, for an even higher number, similar results are to be expected.
Each combination of price signal and aggregation method is repeated ten times.
We use the same set of random number seeds for each combination to ensure comparability between the methods.
Random elements include driving profile generation, selecting a price week for the REAL and FUTURE price signals, and selecting profiles for the \ac{REP} approach.

The performance of each approach is highly dependent on the price signal.
All methods perform significantly worse for the FUTURE price signal.
This is likely because the FUTURE price signal is the most irregular,
with some days almost flat, while others show multiple peaks and valleys.
This increases the complexity of the optimization task and
means that the small errors introduced during aggregation and disaggregation can have a large effect.
That all methods perform worse on the FUTURE price signal is problematic 
because they become most useful in future energy systems with extreme numbers of \acp{EV}.
However, the overall benefit of smart charging is still largely preserved.
All methods perform significantly better than uncontrolled charging.
Uncontrolled charging results in more than double the costs compared to optimal charging;
the different methods only cause additional costs between 15-50 \%.
Note, however, that all methods include an optimization step with perfect foresight.
Therefore, in reality,
forecast errors will come on top of the aggregation error we evaluate in this paper.

The variance between iteration runs seen in \Cref{fig:rsltGap} almost entirely derives from
the fact that we select a random price week for each iteration of the REAL and FUTURE price signals.
If we pick fixed price signals for all iterations but keep the remaining random elements,
the gaps vary at most by 1\% between iteration runs.
The exception is the REP method,
where the random selection of profiles results in \textit{gap} deviations of up to 5\%,
even with a fixed price signal,
because the performance  depends on how well the 500 profiles represent the entire fleet.

\begin{figure}
    \centering
    \includegraphics[width=0.9\columnwidth]{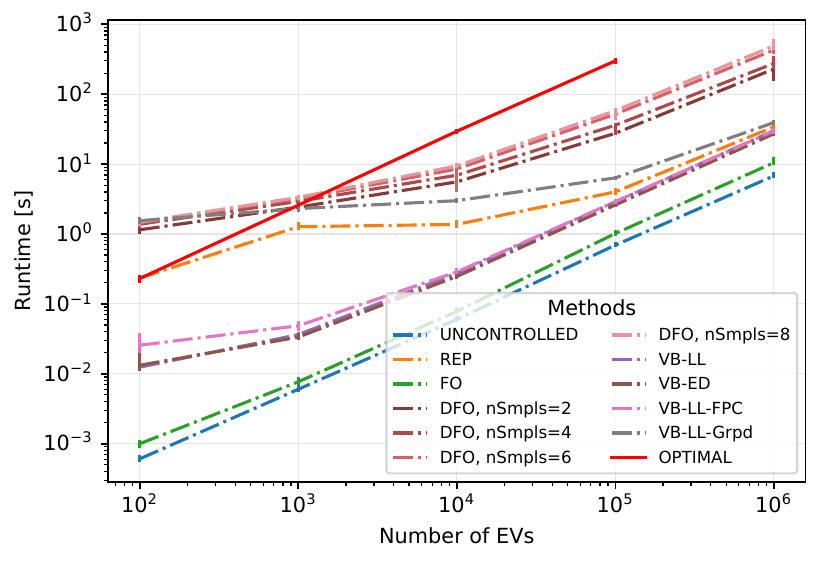}
    \caption{Runtimes of the methods for up to 1,000,000 \acp{EV} (Rust implementation)\protect\footnotemark.
    The OPTIMAL case could not be computed for 1,000,000 because of a lack of memory.
    Runtimes include: creation of flexibility objects from charging events, aggregation, optimizing the aggregated objects, and disaggregation.
    The error bars indicate the minimum and maximum runtime over all runs.
    }
    \label{fig:rsltRT}
\end{figure}

The \ac{FO} method is especially affected by the choice of the price signal.
Compared with the other methods,
\acp{FO} perform very well for the REAL and FUTURE price signals.
However, for the SINE price signal, they perform the worst of all methods.
The good performance for REAL and FUTURE is remarkable because the \ac{FO} method can only shift the charging event's start
but not shape its profile.
Interestingly, in most cases, this limitation is less detrimental than the
\ac{VB}'s overestimation of the flexibility.
However, it appears that in cases of very smooth price signals (like SINE),
this limitation becomes relevant.
The runtime of \acp{FO} is extremely low,
with runtimes comparable to uncontrolled charging,
as seen in \Cref{fig:rsltRT}.

\footnotetext{These results are meant for qualitative comparison of the methods.
Our implementations are as close as possible to the formulations described in the sources.
Therefore, runtime improvements are certainly possible.
For example, we did not exploit parallelization possibilities.}

In terms of \textit{gap}, the \ac{VB} approach performs overall the worst.
In some cases, it can outperforms \acp{FO} and \acp{DFO}.
The runtime is acceptable but can become relevant for larger numbers of \acp{EV}.
The performance of \acp{VB} can be increased at no significant runtime costs
by introducing the piece-wise linear constraint presented in \Cref{sec:vb}.
Grouping the charging events before creating the \ac{VB} also improves performance.
The additional runtime cost of this enhancement is only significant for lower numbers of \acp{EV}.
Overall, the \ac{VB} method works sufficiently well in all aspects
and has the added benefit of being easy to implement and extend (for example, to \ac{V2G}).

\acp{DFO} have reliable good performace for all price signals.
Increasing the $NumSamples$ parameter increases the performance at the cost of additional runtime.
However, the gains diminish strongly after $NumSampels=4$.
This method distinguishes itself with reliability,
especially since no errors occur during disaggregation.
The only other method for which this is the case is \ac{FO}.
However, here, we see stark performance fluctuations with changing price signals.
The disadvantage of this method is its high runtime (See \Cref{fig:rsltRT}).
We see only a speed-up of one order of magnitude compared to the optimal method.
That is with an implementation in Rust.
In a programming language less suited for the many loop-heavy computations required,
like Python, the runtimes exceed those of OPTIMAL\footnote{We provide both Rust and Python implementations on the GitHub page}.
On the other hand, the \ac{DFO} and \ac{FO} approaches have the largest potential for parallelization, which is not utilized in our work.

The representative profile approach (REP) achieves the overall best performance.
Comparatively, the \textit{gap} for REP is very good in all cases.
The runtime is acceptable, comparable to that of \acp{VB}.
Additionally, this method is the only where we do not have to first simulate the total number of \acp{EV} before we can aggregate them.
If we have a stochastic model for \acp{EV}, we can simulate the representative profiles directly without first simulating the entire fleet.
This drastically reduces the overall runtime
but is only applicable in a system modeling context and
only possible if we want to avoid disaggregating the profiles.
However, the method has two disadvantages.
Firstly, the disaggregation is associated with an error.
Secondly, integrating it into a unit commitment model introduces many constraints and variables.
The same is the case for \acp{DFO}.
This point is the strongest advantage of the ungrouped \ac{VB} methods,
where the least additional constraints are introduced to unit commitment.

\section{Sensitivity Analysis}
\label{Sec:sensi}

It is conceivable that the performance of the analyzed methods is strongly dependent on the 
assumptions we make in \Cref{tab:Param}.

To evaluate this, we conduct a sensitivity analysis for the following alternative assumptions.
Each set differs from the base case in \Cref{tab:Param} by one parameter.

\begin{itemize}
    \item WORK: In this scenario, charging is also always possible at the workplace with a charging rate of 11 kW.
    \item WINTER: In winter, the consumption of \acp{EV} can be up to 50\% higher than in summer \cite{argonne}.
    This scenario tests whether this significantly impacts the performance of the methods by increasing the consumption of the average vehicle to 22 kWh/100km.
    \item PHIGH: The charging rate is set to 22 kW.
    \item PLOW: The charging rate is set to 4.7 kW.
\end{itemize}

\begin{figure}
    \centering
    \includegraphics[width=1\columnwidth]{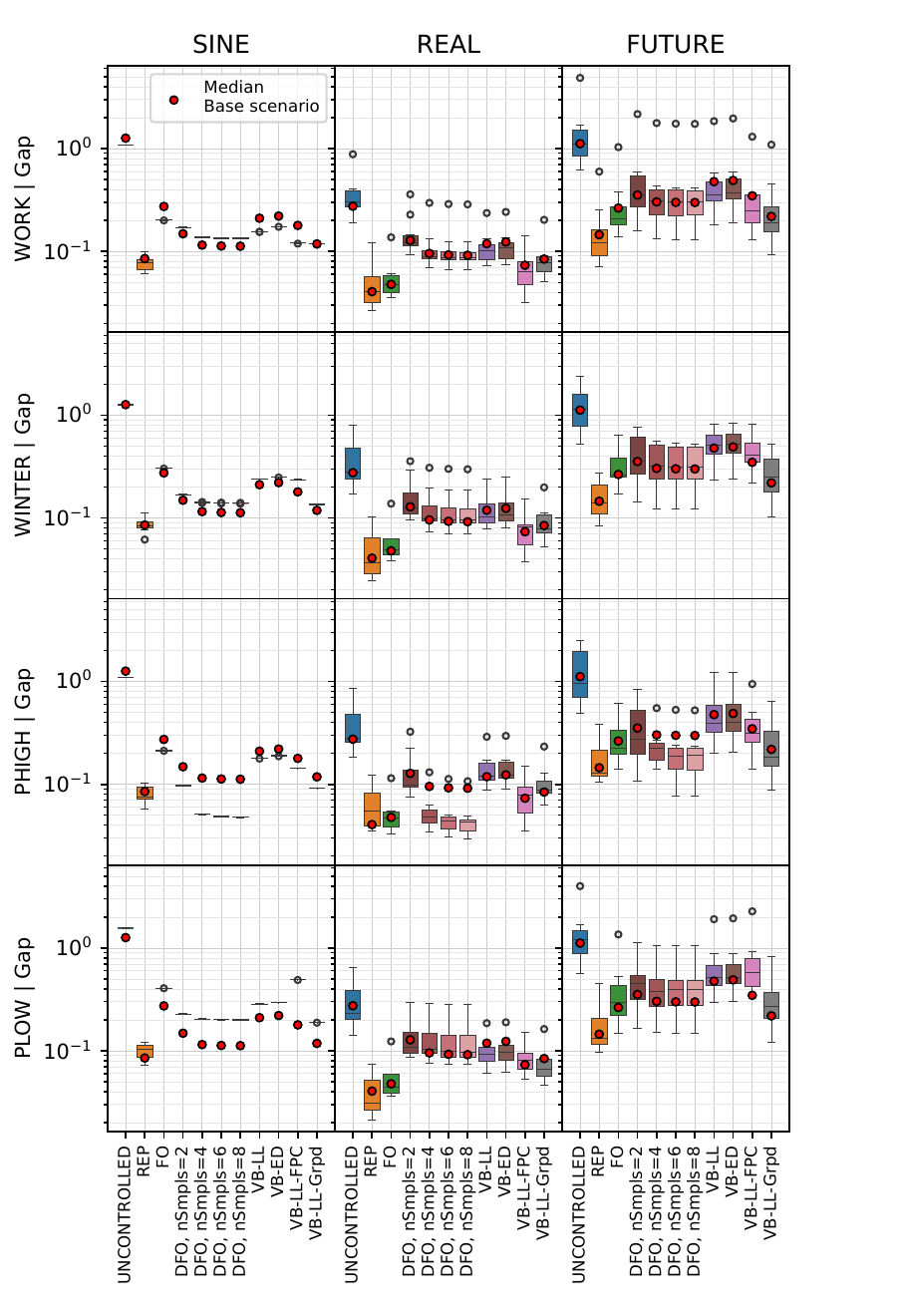}
    \caption{Sensitivity analysis:
    Performance of the different methods on the benchmark for 100,000 \acp{EV}
    for different sets of assumptions.
    The red dots indicate the median gap with the original set of assumptions (\Cref{tab:Param}).}
    \label{fig:rsltSensi}
\end{figure}

In \Cref{fig:rsltSensi},
the results of the sensitivity analysis are summarized.
The WORK and WINTER scenarios change the results only marginally;
this indicates that the performance of the aggregation methods does not strongly depend
on the availability of charging stations at different locations and on the consumption rates of the vehicles.
For the test cases with different charging rates, PHIGH and PLOW, we see minor changes for most methods but substantial differences for the \ac{DFO} method.
The \ac{DFO} method performs significantly worse for lower and better for higher charging rates.
A possible reason for this is that increases in charging rate generally increase the flexibility of charging events.
As stated in \Cref{sec:DFO}, \acp{DFO} perform poorly if slices with no flexibility are aggregated with slices with flexibility.
This problem might extend to slices with very little flexibility.
Therefore, increasing flexibility across the board could lead to fewer problematic cases and better overall performance.

In future work, it will be interesting to analyze whether  the results also remain consistent
for different sets of mobility data,
such as the ACN dataset \cite{leeACNDataAnalysisApplications2019a}.

\section{Conclusion}
\label{sec:conclusion}

In this work, we create a benchmarking setup to evaluate the performance of multiple aggregation-disaggregation pipelines
that enable the integration of large numbers of \acp{EV} into energy system models.
The benchmarking setup samples driving profiles from the \ac{MID} dataset,
assigns \acp{EV} to them,
runs the pipelines,
and evaluates the resulting charging costs for three different price signals.

We tested the commonly used methods of representative profile (REP) and \acf{VB}
as well as \acfp{FO} \cite{siksnysAggregatingDisaggregatingFlexibility2015a} and \acfp{DFO} \cite{siksnysDependencybasedFlexOffersScalable2016}.

For each method, we conduct extensive parameter tuning.

Our results show that no method dominates the others in all aspects.
The \ac{FO} and \ac{DFO} methods have the desirable ability of lossless disaggregation.
This means that any charging load feasible for the aggregated objects is guaranteed to be feasible
for the original set of charging events.
Furthermore, the methods provide algorithms to determine the load of each individual \ac{EV} that will reproduce the aggregated load if summed up.
This aspect is highly desirable in real-world use cases because it guarantees that we can keep any promises we make based on aggregated information.
This advantage comes at the cost of a complex mathematical formulation that is difficult to extend with new features (like \ac{V2G}).
In the case of \ac{DFO}, the runtimes are also significant.
However, both methods perform well in terms of minimizing the charging costs.

The common methods REP and \ac{VB} have the advantage of simplicity.
Our results show that especially the REP method performs very well.
This is encouraging because this method is the most commonly used approach.
\acp{VB} are less performant.
However, they can be improved significantly by introducing an additional piece-wise linear constraint that limits the maximum charging rate
at higher \acp{SOC}.
Grouping the charging events into multiple \acp{VB} also increases the performance.
We tested the least-laxity-first and earliest-departure-first algorithms for disaggregation from \ac{VB} back to individual \acp{EV}.
These disaggregation methods perform approximately the same, with a slight advantage for the least-laxity-first algorithm.
The main advantage of the \ac{VB} method is that it introduces the least amount of constraints and variables
to the optimization that ultimately decides on the best load shape.
This aspect might be desirable if that optimization is already a large unit commitment problem.

Overall, all methods have a clear use case.
We recommend the REP method for scientific energy system modeling when small overestimations of flexibility are acceptable.
Alternatively, if the number of constraints introduced by the REP method is problematic,
we advise to falling back to the \ac{VB} method.
For real-world use cases,
we recommend the \ac{FO} and \ac{DFO} methods.
\ac{FO} if the task has heavy runtime restrictions and \ac{DFO} otherwise.

\section{Open source}
\label{sec:open source}

The benchmarking setup and implementations of the aggregation-disaggregation pipelines are available
on Github: \url{https://github.com/L-Strobel/bench_ev_agg_disagg}

\begin{acks}
This paper is based on research conducted in the ESM-Regio project (https://www.bayerninnovativ.de/de/seite/esm-regio-en)
and is made possible through funding from the German Federal Ministry for Economic Affairs and Climate Action.
\end{acks}

\bibliographystyle{ACM-Reference-Format}
\bibliography{compare-flex.bib}

\appendix

\section{Proof: Flexibility overestimation of Virtual Batteries}
\label{apdx:proof}

We will prove that sufficiently large \acp{VB} overestimate the flexibility of \acp{EV}.

First, we will show no power time series are feasible for the constituent \acp{EV} that violate the constraints of the \ac{VB}.
Therefore, \acp{VB} never underestimate the flexibility of \acp{EV}.

Let $V$ be the set of \acp{EV} that make up the \ac{VB}.
Then:

\begin{equation}
    e_{t}^{VB} = \sum_{v \in V}{e_{t}^{v}} \quad \forall t \in T
    \label{eq:proof1}
\end{equation}

The same is true for $e_{t}^{VB, min}$, $e_{t}^{VB, max}$, $p_{t}^{VB}$, $p_{t}^{VB, min}$, $p_{t}^{VB, max}$, $arr_t$, and $dep_t$.
In other words, all parameters and variables of the individual \acp{EV} have cumulated equivalents in the \ac{VB}.

We can write \Cref{eq:cnstr1} as:

\begin{equation}
    \sum_{v \in V}{e_{t}^{v, min}} \leq \sum_{v \in V}{e_{t}^{v}} \leq \sum_{v \in V}{e_{t}^{v, max}} \quad \forall t \in T
    \label{eq:proof2}
\end{equation}

For every feasible power time series of an individual \ac{EV} $v$ it must be that
$e_{t}^{v, min} \leq e_{t}^{v} \leq e_{t}^{v, max}$ $\forall t \in T$.
If that is the case than \Cref{eq:proof2} is also not violated.

The same logic can be applied to \Cref{eq:cnstr2,eq:cnstr3} if $\Delta t$ and $\eta$ are constants.
Therefore, any power time series that does not violate the constraints of the individual \acp{EV} will also not violate the constraints of the \ac{VB}.

The opposite is not the case. We prove this by counter example.

Let a \ac{VB} be made up of two \acp{EV} A and B.
We look at the time horizon $T=(0, 1, 2, 3)$.
Let $\Delta t = 1$, $\eta = 1$, $e^{v, min}_{t} = 0$, $p^{v, min}_{t} = 0$, $p^{v, max}_{t}= 1$ for $v \in \{A, B\}$ and $\forall t \in T$.
Let $e^{A,max} = (0,1,1,1)$ and $e^{B,max} = (0,1,2,3)$.

This example corresponds to the case where two \acp{EV} are plugged in simultaneously.
Both don't need any electricity to drive their upcoming trips.
Both are not fully charged; B can charge three times more than A.

At t=0 the \ac{VB} will not charge.
In that case,
at t=1 the \ac{VB} can charge with $p^{VB}_{1}=2$,
because:

\begin{equation}
    e^{VB}_{0} = e^{VB}_{1} = 0
    \label{eq:proof3}
\end{equation}

\begin{equation}
    p^{VB, max}_{t} = p^{A, max}_{t} +  p^{B, max}_{t} = 2 \quad \forall t \in T
    \label{eq:proof4}
\end{equation}

\begin{equation}
    e^{VB, max}_{2} = e^{A, max}_{2} +  e^{B, max}_{2} = 3.
    \label{eq:proof5}
\end{equation}

In that case, it must be that $p^{A}_{1} = p^{A, max}_{1} = 1$ and, consequently, $e^{A}_{2} = e^{A, max}_{3} = 1$.
I.e., A must be fully charged after t=1.

Similarly, the \ac{VB} can charge with $p^{VB}_{2}=2$ in t=2,
because of \Cref{eq:proof4} and:

\begin{equation}
     e^{VB}_{2} = e^{VB}_{1} + p^{VB}_{1} \cdot \Delta t \cdot \eta = 2
    \label{eq:proof6}
\end{equation}

\begin{equation}
    e^{VB, max}_{3} = e^{A, max}_{3} +  e^{B, max}_{3} = 4
    \label{eq:proof7}
\end{equation}

However, this also necessitates that A charges, but A is already fully charged.
Therefore, this power time series ($p^{VB}=(0,2,2,0)$), which is possible for the \ac{VB},
violates the constraints of an individual \ac{EV}.

The example above can be generalized.
It is only necessary that both \acp{EV} are plugged in simultaneously for longer than the A needs to fully charge plus one time step.
Furthermore, the difference between the energy B and A can charge must be greater than that B can charge in one time step.

The overestimation effect of situations such as the example above is never compensated
because \acp{VB} never underestimate the flexibility of \acp{EV}.
Consequently, whenever the probability of these situations is none-zero (which, judging from experience, is always the case),
any \ac{VB} consisting of a sufficient number of \acp{EV} will overestimate their flexibility.

\section{OPTIMAL}
\label{apdx:opt}

The true optimal charging costs are determined with the following optimization:

\begin{equation}
    \underset{p_{t, v}}{minimize} \sum_{t \in T}{\sum_{v \in V}{p_{t, v} \cdot \Delta t \cdot c_{t}}}
    \label{eq:o_costs}
\end{equation}

subject to:

\begin{equation}
    0 \leq p_{t, v} \leq \overline{p}_{t,v} \quad \forall t \in T, \forall v \in V
    \label{eq:o_cnstr1}
\end{equation}

\begin{equation}
    0 \leq e_{t, v} \leq \overline{e}_{t,v} \quad \forall t \in T, \forall v \in V
    \label{eq:o_cnstr2}
\end{equation}

\begin{equation}
    e_{\hat{t}, v} = u_{\hat{t}, v} \quad \forall \hat{t} \in S, \forall v \in V
    \label{eq:o_cnstr3}
\end{equation}

\begin{equation}
    e_{t+1, v} = e_{t, v} + p_{t, v} \cdot \Delta t \cdot \eta - cons_{t} \quad \forall t \in T, \forall v \in V
    \label{eq:o_cnstr4}
\end{equation}

$p_{t, v}$ is the charging power of \ac{EV} $v$ at time $t$.
$c_t$ are the charging costs at time $t$.
$T$ is the set of relevant time steps.
$V$ is the set of \acp{EV}.
\Cref{eq:o_cnstr1} defines the minimum and maximum charging power ($\overline{p}_{t,v}$) of each \ac{EV}.
Note, that $\overline{p}_{t,v}$ is zero if the vehicle is not plugged in.
\Cref{eq:o_cnstr2} defines the minimum and maximum battery energy content ($\overline{e}_{t,v}$).
\Cref{eq:o_cnstr3} sets the energy content at the start and end of each charging event equal to that in the simulation of uncontrolled charging.
$u_{\hat{t}, v}$ is the energy content at time $\hat{t}$ in the simulation of uncontrolled charging.
Here $\hat{t}$, is a start or end time step of a charging event, defined by the set $S$.
\Cref{eq:o_cnstr4} defines the charging and consumption mechanisms.

The \ac{EV} parameters assumptions and generated driving profiles provide the necessary data
for the plug-in time windows, consumption time series, maximum charging power, battery capacity, vehicle consumption, etc.  (See \Cref{sec:setup}).

We implemented the optimization in Python using \textit{gurobipy} and the Gurobi 11.0 solver.

\begin{acronym}
	\acro{EV}{electric vehicle}
    \acro{REP}{representative profile}
	\acro{EVSE}{electric vehicle supply equipment}
	\acro{SOC}{state-of-charge}
    \acro{FO}{Flex Object}
    \acro{DFO}{Dependency-based FlexOffer}
    \acro{MID}{"Mobilität in Deutschland 2017"}
    \acro{V2G}{vehilce-to-grid}
    \acro{VB}{virtual battery}
\end{acronym}

\end{document}